\documentclass[twocolumn,twoside,slac_two]{revtex4}
\usepackage{graphicx}
\usepackage{fancyhdr}
\begin{document}

\title{A diagnostic test for determining the location of the GeV emission in powerful blazars}

\author{Amanda Dotson}
\affiliation{Department of Physics, Joint Center for Astrophysics,
University of Maryland Baltimore County, 1000 Hilltop Circle,
Baltimore, MD 21250, USA}

\author{Markos Georganopoulos}
\affiliation{Department of Physics, Joint Center for Astrophysics, 
University of Maryland Baltimore County, 1000 Hilltop Circle, Baltimore, MD 21250, USA
\\
NASA Goddard Space Flight Center, Code 660, Greenbelt, MD 20771, USA}

\author{Demosthenes Kazanas}
\affiliation{NASA Goddard Space
Flight Center, Code 660, Greenbelt, MD 20771, USA}

\author{Eric Perlman}
\affiliation{Department of Physics and Space Sciences,
Florida Institute of Technology, 150 West University Boulevard,
Melbourne, FL  32901, USA}

\begin{abstract}
 	An issue currently under debate in the literature is how far from the black hole is the {\sl Fermi}-observed GeV emission of powerful blazars emitted. Here we present a clear diagnostic tool for testing whether the GeV emission site is located within the sub-pc broad emission line (BLR) region or further out in the few pc scale molecular torus (MT) environment.  Within the BLR the scattering takes place at the onset of the Klein-Nishina regime, causing the electron cooling time to become almost energy independent and as a result, the variation of high-energy emission is expected to be achromatic.  Contrarily, if the emission site is located outside the BLR, the expected GeV variability is energy-dependent and with amplitude increasing with energy.  We demonstrate this using time-dependent numerical simulations of blazar variability.  
\end{abstract}

\maketitle

\thispagestyle{fancy}

\section{Introduction}
\label{sec:intro}
	Blazars are by far the most common objects detected in the gamma-ray sky \citep{abdo112fgl}. \textit{Fermi} has detected blazar variability as short as a few hours \citep[e.g.][]{abdo10b}. During these flares, the GeV luminosity has been known to increase by a factor of up to several compared to its pre-flare luminosity.  Because blazars cannot be resolved at these energies (or at any other energy, with the possible exception of VLBA observations), it is impossible to determine the location of these flares by direct detection.

	To address this issue, we propose a diagnostic test that utilizes \textit{Fermi} variability data of short flares to determine the location of the GeV emission in blazars.  
	
\section{Sources of Seed Photons}	
	
	 Relativistic effects determine which photon field is dominant at varying distances from the central black hole, and as a result the location of the GeV flaring site determines the dominant source of seed photons.  The co-moving (jet frame) energy density of a radiation field $U'$ scales as differing factors of $\Gamma$ depending on the direction from which the photons enter the emitting region \citep{dermer94,geo01}. If the photon field is isotropic $U' \approx \Gamma^2 U$. For photons entering the emitting region from behind the relativistically moving blob $U' \approx  U\Gamma^{-2}$.  If we assume a nominal FSRQ accretion disk luminosity of $L_{disk}\sim10^{45}$ erg s$^{-1}$ and that a fraction $\xi = 0.1$ of this radiation is reprocessed by  both the BLR and the MT, a typical luminosity of the external radiation field is $L_{ext} \sim 10^{44}$ erg s$^{-1}$ \citep{ghisellini09}.

	If the emission site is located within the BLR (at $R \sim 10^{17} \mathrm{cm}$), the photon field can be considered isotropic in the galaxy frame and its co-moving energy density is $U'_{BLR} \approx \,2.6 \; \Gamma_{10}^{2} L_{BLR, 44} R^{-2}_{BLR,17} \;  \mbox{erg cm}^{-3}$. Similarly, the  MT photon field is isotropic inside the BLR and its co-moving seed photon energy density is $U'_{MT}\approx  2.6 \times 10^{-2} \; \Gamma_{10}^2 L_{MT, 44} R^{-2}_{MT,18} \;  \mbox{erg cm}^{-3}$. Clearly, inside the $R_{BLR}$ the co-moving BLR photon field energy density  dominates over that of the MT by a factor of $\sim 100$. 
 
	 If the emission site is located at $R \sim 10^{18} \mathrm{cm} $ (within the MT) then the BLR  UV photons enter the emitting region practically from behind, so that $U'_{BLR}  \approx 2.6 \times 10^{-4} \; L_{BLR, 44} R^{-2}_{MT,18} \Gamma_{10}^{-2}\; \mbox{erg cm}^{-3}$.  The IR photons from the MT  retain the same co-moving energy density previously given by $U'_{MT}\approx  2.6 \times 10^{-2} \; \Gamma_{10}^2 L_{MT, 44} R^{-2}_{MT,18} \;  \mbox{erg cm}^{-3}$.  In this case, therefore, it is the MT  that dominates the co-moving photon energy density.
 
 	These external photon field co-moving luminosities need to be compared to the synchrotron photon field energy density. If $R_{blob}$ is the size of the emitting blob, the co-moving synchrotron photon energy density is $U'_{S}  \approx { L_{S}\over 4\pi c R_{blob}^2 \Gamma^4}$.  The most plausible assumption for the size of the emitting region, however, is to set an upper limit to it by its variability timescale: $R_{blob}=c t_{var} \delta $.  We then obtain a lower limit for the synchrotron energy density $U'_{S}  \approx 6.3 \times 10^{-2} L_{46} t_{var, 6h}^{-2} \Gamma_{10}^{-6} \;  \mbox{erg cm}^{-3}$, where we used a 6 hour variability scale, as seen in variability timescales observed by Fermi \citep{abdo10b}.   
  
  	Although the synchrotron photon energy density is  substantially lower than the BLR photon energy density when the blazar emission site is within the BLR, it is comparable or even higher than the MT photon energy density when the blazar emission site is within the MT but outside the BLR.  We briefly note here if we want to have a blob of a given size $R_{blob}$ at a distance larger than $R/\Gamma$, \citep[e.g.][]{marscher08} the blob cannot occupy the entire cross section of the jet.  Our diagnostic hinges on the fact that if the blazar emission site is located outside the BLR, electron cooling occurs in the Thomson regime.  This is always true for the case of the MT photons being dominant and it can be shown that if SSC photons dominate, cooling still takes place within the Thomson regime for powerful FSRQs.
 
 \section{Cooling in the BLR vs Cooling in the MT}
	{\sl The critical difference between the BLR and the MT is the energy of the seed photons: photons originating from the BLR are UV photons ($\epsilon_0 \approx 10^{-5}$) while photons originating from the MT are IR photons ($\epsilon_0 \approx 10^{-7}$).  This difference by a factor of $\sim 10^2$ in typical photon energy is critical in that it affects the energy regime in which electron cooling takes place, and thus the energy dependence of the electron cooling time.}

In powerful FSRQs the IC emission in high states can dominate over the synchrotron emission by a factor of up to $\sim100$ \citep[e.g.][]{abdo10a}. In cases of high Compton dominance, the primary electron cooling mechanism is IC scattering.  For electrons cooling in the Thomson regime ($\gamma\epsilon_0 \lesssim 1$, where both the electron Lorentz factor $\gamma$ and the seed photon energy $\epsilon_0$ are measured in the same frame) the cooling rate $\dot{\gamma} \propto \gamma^2$.  For electrons with $\gamma \epsilon_0 \gtrsim 1$ cooling takes place in the Klein-Nishina (KN) regime with $\dot{\gamma} \propto \ln \gamma$ \citep{blum70}.  Given that the BLR photons are $\sim 10^2$ more energetic than the MT photons, we expect that if cooling takes place in the BLR, effects of the transition from the Thomson to the KN regime will be manifested in electrons of  energy $\sim 10^{-2}$  times lower than if the cooling takes place within the MT.  
\begin{figure}
	
\includegraphics[width=65mm]{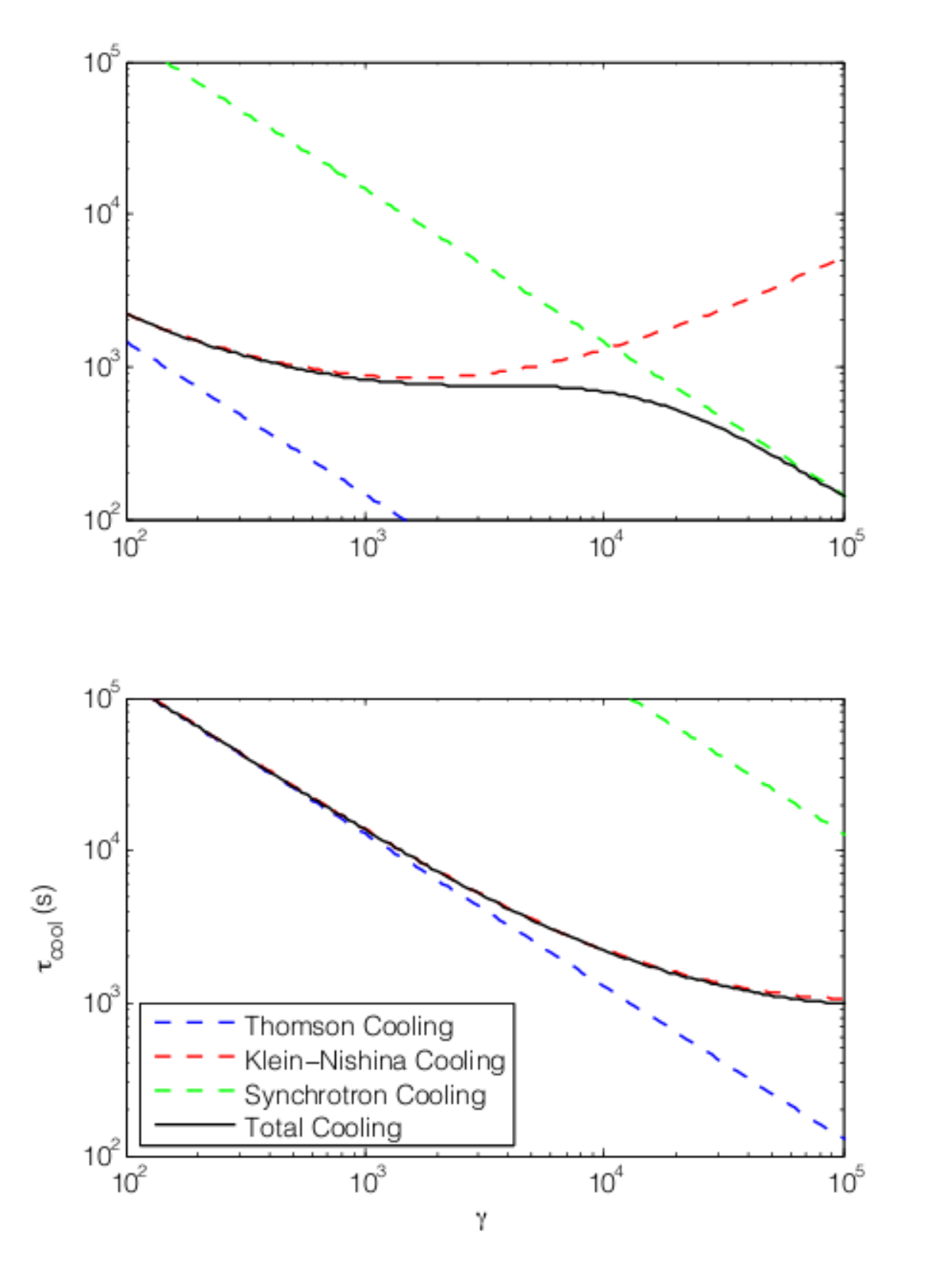}

\caption{Cooling time in the blob frame as a function of co-moving electron energy $\gamma$. Top panel: Blazar emission site located in the BLR. Bottom panel: Blazar emission site located in the MT.  The dotted lines represent the various cooling mechanisms (blue = Thomson cooling, red = KN cooling, green = synchrotron cooling), the solid black line is the total cooling time.  Plots were calculated for the following values: seed photon energies: $\epsilon_{0,BLR} = 3 \times 10^{-6}$ and $\epsilon_{0,MT} = 6 \times 10^{-7}$, energy densities $U_{BLR} = 2.65 \times 10^{-2}$ erg cm$^{-3}$, $U_{MT} = 3 \times 10^{-4}$ erg cm$^{-3}$, a $U_B$ that corresponded to a Compton dominance of $\sim 100$, and $\Gamma_{bulk} = 20$.}

\label{fig:tblrmt}
\end{figure}

	The effects of the transition between Thomson and KN regimes on the electron energy distribution (EED) and the resultant spectrum of the synchrotron and IC emission have been studied before \citep[e.g.][]{blum71,zdz89,dermer93,sok04,mod05,kusun05,georganopoulos06,manolakou07,sikora09}.  In short, because $\dot{\gamma} \propto \gamma^2$ in the Thomson regime and $\dot{\gamma} \propto \ln\gamma$ in the KN regime, the cooling time $\tau_{cool} = \gamma/\dot{\gamma}$ scales as $\gamma^{-1}$ in the Thomson regime and as $\gamma/\ln\gamma$ in the KN regime.  

	 Because  the cooling time  is approximately energy independent around $\gamma \epsilon_0 \sim 1$, this energy-independent cooling time will be manifested at energies lower by a factor of  $\sim 100$  for cooling taking place in the BLR compared to cooling taking place  in the MT, since the BLR seed photons have an energy higher than that of the MT by a factor of $\sim 100$. This can be seen in Fig. \ref{fig:tblrmt} where we plot the electron cooling time for a source with a ratio of external photon field energy density  $U_0'$ in the co-moving frame to co-moving magnetic field energy density $U_B$,  $U_0'/U_B=100$. This corresponds to a factor of $\sim 100$ Compton dominance (ratio of inverse Compton to Synchrotron luminosity), similar to what is observed in the most Compton dominated sources.

	The transition from Thomson cooling to KN cooling, and from KN cooling to synchrotron cooling can also be seen in Fig. \ref{fig:tblrmt}.  If the electrons are cooling on photons from the BLR, cooling takes place in the Klein-Nishina regime and the cooling time scale is approximately energy independent around $\gamma \epsilon_0 \sim 1$ (Fig. \ref{fig:tblrmt}, top panel).  If the electron population cools on photons from the molecular torus, cooling takes place in the Thomson regime (Fig. \ref{fig:tblrmt}, bottom panel).  The cooling time is heavily energy dependent, and any variations should consequently exhibit heavy energy-dependence.  

\section{The Diagnostic Test}
	The energy dependence of the cooling time results in an energy dependence (or lack thereof) of variations: if the blazar emission site is located within the BLR, variations should be achromatic.  The energy dependence of the variations can be used as a diagnostic test to determine if the GeV emission site is located within the BLR.  By comparing \textit{Fermi} light curves of flares at different energies, we propose that the energy dependence of the light curve can be used as a diagnostic test to rule out whether the GeV flare originates in the BLR or MT.

\subsection{Numerical Simulation Results}
	To demonstrate the effect of the energy independence or dependence of the electron cooling time on the variability of a flare, we utilized a one-zone numerical model to simulate a flare. We initialized the code with values appropriate for a high power blazar with a Compton dominance of $\sim 100$.  For this particular simulation we assumed a source size $R=10^{16} \mathrm{cm}$, bulk Lorentz factor $\Gamma = 10$, co-moving injected electron luminosity $L = 2 \times 10^{44}$ erg s $^{-1}$, maximum electron Lorentz factor $\gamma_{max} = 10^5$, and electron index $p=2.5$.  For the case of a flaring region located within the BLR we assumed an initial photon energy $\epsilon_0 = 3 \times 10^{-5}$ and an energy density (in the galaxy frame) $U_{BLR}=2.6 \times 10^{-2}$ erg cm$^{-3}$.  For the case of a flaring region located outside the BLR we assumed an initial photon energy $\epsilon_0 = 6 \times 10^{-7}$ and an energy density $U_{MT}=3 \times 10^{-4}$ erg cm$^{-3}$.  For each case the magnetic field \textbf{B} was fixed to assume a Compton dominance $U_{EC}/U_B=100$.
	
\begin{figure}
	\includegraphics[width=65mm]{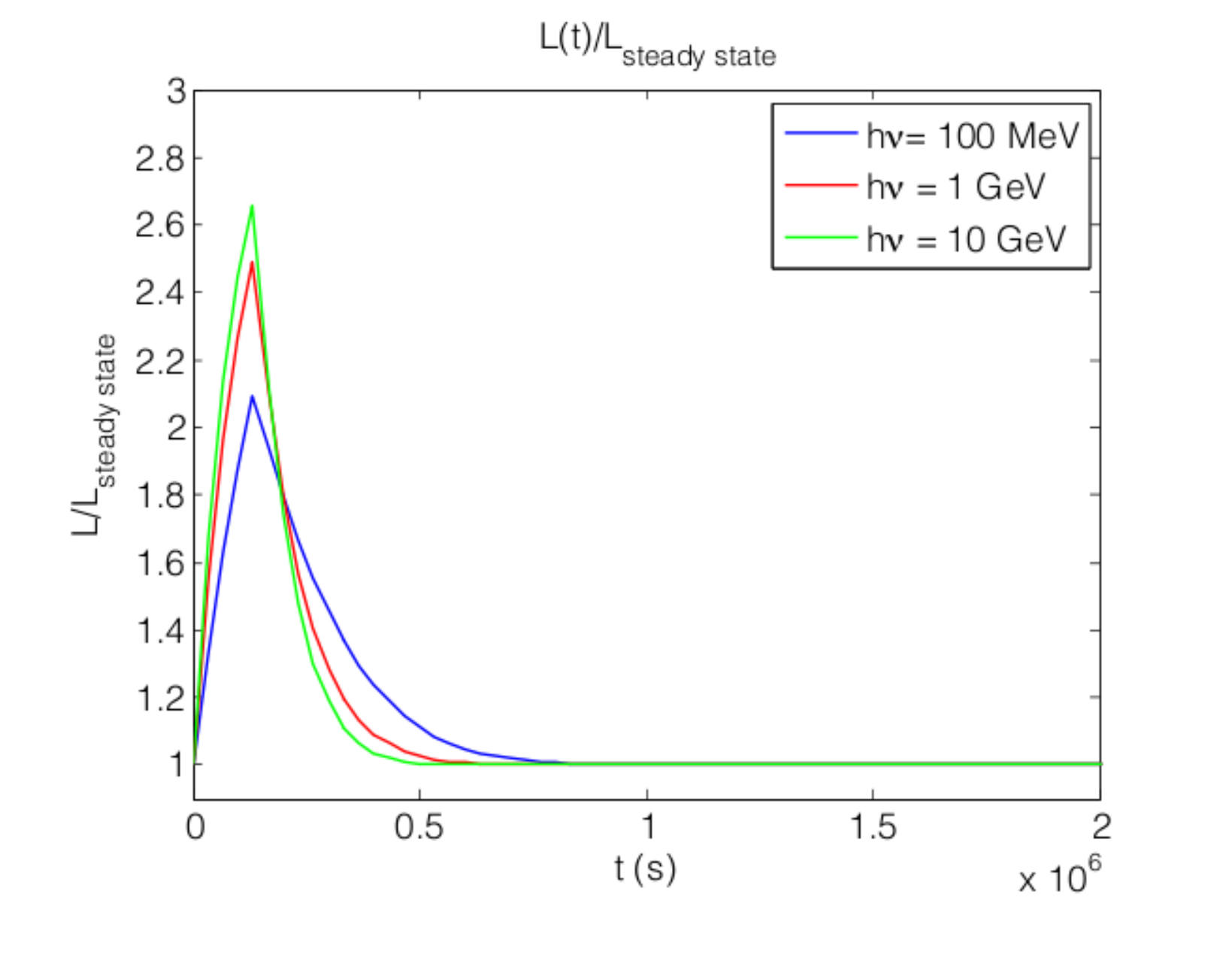}
	\caption{Light curve at various energies (BLR seed photons): $\epsilon_0=100$ MeV, $\epsilon_0=1$ GeV , 
	$\epsilon_0=10$ GeV. }
	\label{fig:blrflare}
\end{figure}

\begin{figure}
	\includegraphics[width=65mm]{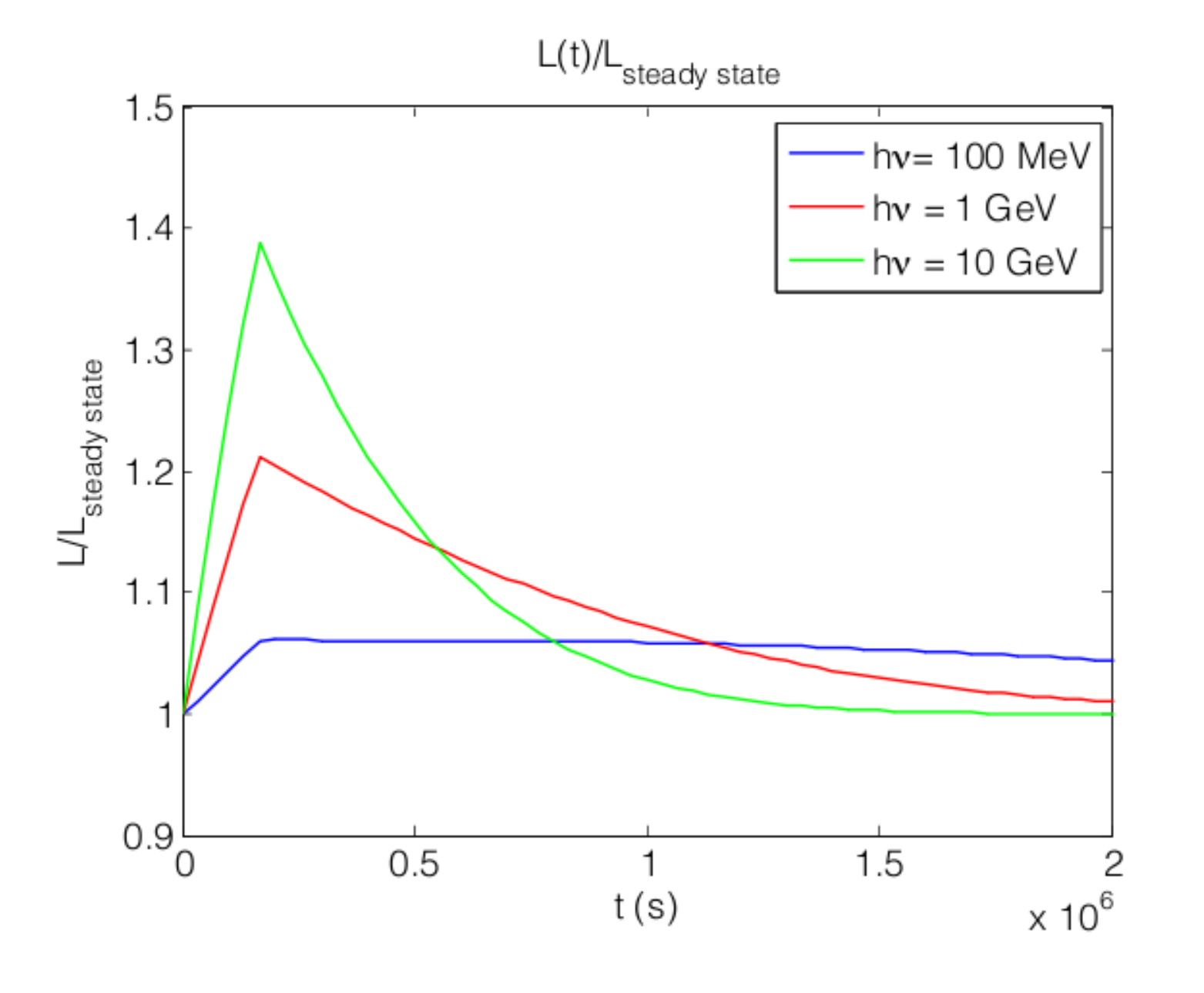}
	\caption{Light curve at various energies (MT seed photons): $\epsilon_0=100$ MeV, $\epsilon_0=1$ GeV, $\epsilon_0=10$ GeV. }
	\label{fig:mflare}
\end{figure}

	The system behaves as expected, showing a noticeable difference in the decay rate as well as the amplitude of the flare depending on if the seed photons originated from within or from outside the BLR (see Figs. \ref{fig:blrflare} and \ref{fig:mflare}).  This difference in decay rate and amplitude can be used as a diagnostic test to differentiate between flares that take place inside or outside the BLR.  

\subsection{Light-Crossing Time Effects}
	Because the GeV emitting region is not a point source, any change in the luminosity of the blob will not be seen instantaneously.  Instead, the observed decay time of the light curve is the result of the actual decay time and the light-crossing time inherent in the blob.  To test whether the light-crossing time would erase any difference in the light curves (as predicted by our diagnostic) for the case of the emitting region being located outside the BLR, we modeled the flux of a source with light-crossing time $t_{LC}$ that is decreasing in flux.  We assumed the flux of the source is decaying exponentially, $F(t) = F_0 \exp^{-t/t_c}$, where $t_c$ is the cooling time at a specific energy.  
	
	Our diagnostic predicts differences in the decay time of the light curves for the case where the emitting region is located within the MT; for the purposes of this demonstration we assume cooling occurs in the Thomson regime and as a result $t_c \propto \epsilon^{-1/2}$.  We plot the resultant light curves for cooling times differing by a factor of $1/\sqrt{10}$ (i.e. energies differing by a factor of 10).
\begin{figure}
	\includegraphics[width=65mm]{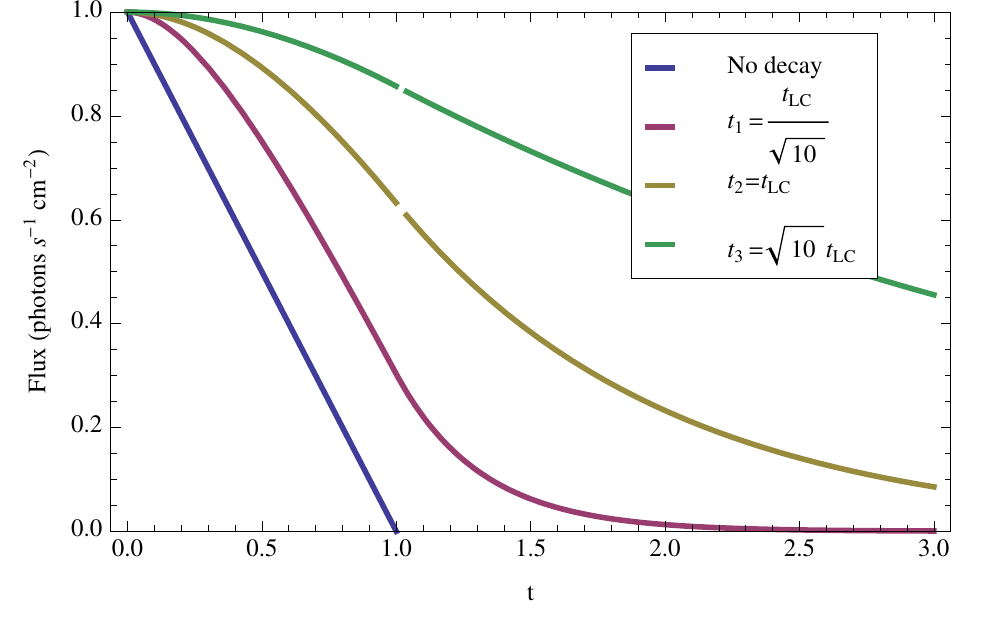}
	\caption{Light curves with light-crossing time effects factored in.  Initial flux and light-crossing time are normalized to 1. Light curves are plotted for energies differing by a factor of 10.}
	\label{fig:lctime}
\end{figure}
As evident in Fig. \ref{fig:lctime}, even with light-travel time effects convolved with exponential decay times, the predicted differences in the decay times are still preserved.  

\subsection{Feasibility Study}
		A flare that occurs outside  the BLR should exhibit energy dependent cooling  times.  Practical application of the diagnostic hinges on the requirement that the decay times in different energy bands should have  detectable differences.  To demonstrate this, we took as a test case a flare of 3C 454.3 \citep{abdo3c454}  and assume an exponential decay of the light-curve $F(\epsilon,t) = N_{b} + N_{0} e^{-t/t_c}$, where $N_b$ and $N_0$ are normalizing factors and $t_c$ is the energy-dependent cooling time. We assumed that the flaring site is located outside the BLR (i.e. cooling occurs in the Thomson regime) because this is case where different decay times need to be resolved.  We created simulated data points at 6-hour time intervals, applied the energy-dependent maximum error from the \textit{Fermi} observations of the flare, and then fit a curve to the simulated data to obtain a cooling time.  From this test, we found that even within maximum observational error, the decay times at different energies are still distinguishable. 
	
\section{Conclusions}
	We have presented a diagnostic test that utilizes blazar variability to determine the location of the GeV emitting site in blazars.  The energy difference in seed photons originating from the BLR versus seed photons originating from the MT causes electrons within the emitting site to cool in different energy regimes.  
	
	For the case where the GeV emitting site is located within the BLR, cooling takes place at the onset of the KN regime, and the resultant electron cooling time is energy-independent.  We have demonstrated that the associated light curves exhibits decay times that are approximately energy independent.  Conversely, for the case where the GeV emitting site is located outside the BLR, cooling takes place in the Thomson regime and the electron cooling times are heavily energy dependent.  In this case, the associated light curves exhibit energy dependence of their decay times.  
	
	The energy dependence of the decay time of the light curves is visible within the \textit{Fermi} energies; these differences can be used as a diagnostic test to determine whether the GeV emitting region is located inside or outside the BLR.  These effects are observable within the maximum measured error of \textit{Fermi} observations and are not erased due to considering light-travel time effects.  If light curves from a sufficiently bright and rapid flare \cite[such as that in 3C 454.4;][]{abdo3c454} are compared at different energies,  if the GeV emitting site is located within the BLR,  the decay times will exhibit no energy dependence, whereas if the emitting site is located within the MT, the decay times will exhibit energy dependence.

\end{document}